\documentclass[twocolumn]{emulateapj}

\newcommand{\etal}{{\it et al.}}
\newcommand {\lsim}{\mbox{$\:\stackrel{<}{_{\sim}}\:$} }
\newcommand {\gsim}{\mbox{$\:\stackrel{>}{_{\sim}}\:$} }

\begin{document}
\title{How Dry is the Brown Dwarf Desert?: Quantifying the Relative Number of Planets, Brown Dwarfs 
and Stellar Companions around Nearby Sun-like Stars}
\medskip
\author{Daniel Grether$^{1}$ \& Charles H. Lineweaver$^{1,2}$}
\affil{$^{1}$ Department of Astrophysics, School of Physics, University of New South Wales,\\
Sydney, NSW 2052, Australia\\}
\affil{$^{2}$ Planetary Science Institute, Research School of Astronomy and Astrophysics \& \\ 
Research School of Earth Sciences, Australian National University, Canberra, ACT, Australia\\}


\begin{abstract}

Sun-like stars have stellar, brown dwarf and planetary companions. To help constrain their formation and migration 
scenarios, we analyse the close companions (orbital period $< 5$ years) of nearby Sun-like stars. By using the same sample 
to extract the relative numbers of stellar, brown dwarf and planetary companions, we verify the existence of a very dry 
brown dwarf desert and describe it quantitatively. With decreasing mass, the companion mass function drops by almost two 
orders of magnitude from $1 \; M_{\odot}$ stellar companions to the brown dwarf desert and then rises by more than an 
order of magnitude from brown dwarfs to Jupiter-mass planets. The slopes of the planetary and stellar companion mass 
functions are of opposite sign and are incompatible at the 3 sigma level, thus yielding a brown dwarf desert. The minimum 
number of companions per unit interval in log mass (the driest part of the desert) is at $M = 31 \:^{+25}_{-18} \: M_{Jup}$. 
Approximately $16\%$ of Sun-like stars have close ($P < 5$ years) companions more massive than Jupiter: $11\% \pm 3\%$ are 
stellar, $<1\%$ are brown dwarf and $5\% \pm 2\%$ are giant planets. The steep decline in the number of companions in the 
brown dwarf regime, compared to the initial mass function of individual stars and free-floating brown dwarfs, suggests 
either a different spectrum of gravitational fragmentation in the formation environment or post-formation migratory 
processes disinclined to leave brown dwarfs in close orbits.
\end{abstract}


\keywords{}


\section{Introduction}

The formation of a binary star via molecular cloud fragmentation and collapse, and the formation of a massive planet 
via accretion around a core in a protoplanetary disk both involve the production of a binary system, but are usually 
recognized as distinct processes (e.g. Heacox 1999; Kroupa \& Bouvier 2003, see however Boss 2002). The formation of 
companion brown dwarfs, with masses in between the stellar and planetary mass ranges, may have elements of both or some 
new mechanism \citep{Bate00, Rice03, Jiang04}. For the purposes of our analysis brown dwarfs can be conveniently defined 
as bodies massive enough to burn deuterium ($M \gsim 13 \: M_{Jup}$), but not massive enough to burn hydrogen 
($M \lsim 80 \: M_{Jup}$ e.g. Burrows 1997). Since fusion does not turn on in gravitationally collapsing 
fragments of a molecular cloud until the final masses of the fragments are largely in place, gravitational collapse, 
fragmentation and accretion should produce a spectrum of masses that does not know about these deuterium and hydrogen 
burning boundaries. Thus, these mass boundaries should not necessarily correspond to transitions in the mode of formation. 
The physics of gravitational collapse, fragmentation, accretion disk stability and the transfer of angular momentum, 
should be responsible for the relative abundances of objects of different masses, not fusion onset limits.

However, there seems to be a brown dwarf desert -- a deficit in the frequency of brown dwarf companions either relative to 
the frequency of less massive planetary companions \citep{Marcy00} or relative to the frequency of more massive stellar 
companions to Sun-like hosts. The goal of this work is (i) to verify that this desert is not a selection effect due to our 
inablility to detect brown dwarfs and (ii) to quantify the brown dwarf desert more carefully with respect to both stars 
and planets. By selecting a single sample of nearby stars as potential hosts for all types of companions, we can better 
control selection effects and more accurately determine the relative number of companions more and less massive than 
brown dwarfs.

Various models have been suggested for the formation of companion stars, brown dwarfs and planets (e.g. Larson 2003,   
Kroupa \& Bouvier 2003, Bate 2000, Matzner \& Levin 2004, Boss 2002, Rice \textit{et al.} 2003). All models involve 
gravitational collapse and a mechanism for the transfer of energy and angular momentum away from the collapsing material. 

\begin{figure}[!t]
\plotone{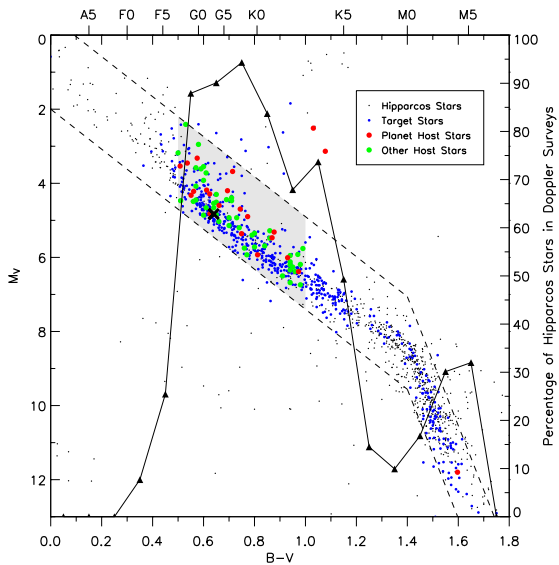}
\scriptsize
\caption{Our Close Sample. Hertzsprung-Russell diagram for Hipparcos stars closer than 25 pc. Small black dots are 
Hipparcos stars not being monitored for possible companions by one of the 8 high precision Doppler surveys considered here 
\citep{Lineweaver03}. Larger blue dots are the subset of Hipparcos stars that are being monitored (``Target Stars") but 
have as yet no known planetary companions. The still larger red dots are the subset of target stars hosting detected 
planets (``Planet Host Stars") and the green dots are those hosts with larger mass ($M_2 > 13 M_{Jup}$) companions 
(``Other Host Stars"). Only companions in our less-biased sample ($P < 5$ years and  $M_2 > 10^{-3} M_\odot$) are shown 
(see Section 2.2). Our Sun is shown as the black cross. The grey parallelogram is the region of $M_v$ - $(B-V)$ space 
that contains the highest fraction (as shown by the triangles) of Hipparcos stars that are being monitored for exoplanets. 
This Sun-like region -- late F to early K type main sequence stars -- contains our Hipparcos Sun-like Stars. The target 
fraction needs to be as high as possible to minimize selection effects potentially associated with companion frequency.
The target fraction is calculated from the number of main sequence stars, i.e., the number of stars in each bin between the 
two dashed lines. This plot contains 1509 Hipparcos stars, of which 627 are Doppler target stars. The Sun-like region 
contains 464 Hipparcos stars, of which 384 are target stars. Thus, the target fraction in the Sun-like grey parallelogram 
is $\sim 83\% (=384/464)$.
}
\label{fig:HR_25}
\end{figure}

\begin{figure}[!t]
\plotone{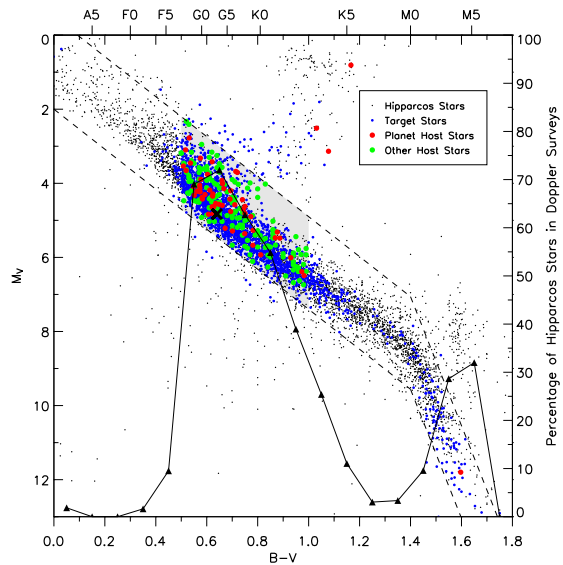}
\caption{Our Far Sample. Same as Fig. \ref{fig:HR_25} but for all Hipparcos stars closer than 50 pc. The major reason the 
target fraction ($\sim 61\%$, triangles) is lower than in the 25 pc sample ($\sim 83\%$) is that K stars become too faint 
to include in many of the high precision Doppler surveys where the apparent magnitude is limited to $V < 7.5$ 
\citep{Lineweaver03}. This plot contains 6924 Hipparcos stars, of which 2351 are target stars. The grey parallelogram 
contains 3296 Hipparcos stars, of which 2001 are high precision Doppler target stars ($61\% \sim 2001/3296$). The stars 
below the main sequence and the stars to the right of the M dwarfs are largely due to uncertainties in the Hipparcos 
parallax or $B-V$ determinations.
}
\label{fig:HR_50}
\end{figure}

Observations of giant planets in close orbits have challenged the conventional view in which giant planets form beyond the 
ice zone and stay there (e.g. Udry 2003). Various types of migration have been proposed to meet this challenge. The most 
important factors in determining the result of the migration is the time of formation and mass of the secondary and its 
relation to the mass and time evolution of the disk (e.g. Armitage \& Bonnell 2002). We may be able to constrain the above 
models by quantitative analysis of the brown dwarf desert. For example, if two distinct processes are responsible for the 
formation of stellar and planetary secondaries, we would expect well-defined slopes of the mass function in these mass 
ranges to meet in a sharp brown dwarf valley.

We examine the mass, and period distributions for companion brown dwarfs and compare them with those of companion stars and 
planets. The work most similar to our analysis has been carried out by \citet{Heacox99, Zucker01b} and \citet{Mazeh03}. 
\citet{Heacox99} and \citet{Zucker01b} both combined the stellar sample of \citet{Duquennoy91} along with the known 
substellar companions and identified different mass functions for the planetary mass regime below 10 $M_{Jup}$ but found 
similar flat distributions in logarithmic mass for brown dwarf and stellar companions. \citet{Heacox99} found that the 
logarithmic mass function in the planetary regime is best fit by a power-law with a slightly negative slope whereas 
\citet{Zucker01b} found an approximately flat distribution. \citet{Mazeh03} looked at a sample of main sequence stars using 
infrared spectroscopy and combined them with the known substellar companions and found that in log mass, the stellar 
companions reduce in number towards the brown dwarf mass range. They identify a flat distribution for planetary mass 
companions. We discuss the comparison of our results to these in Section 3.1.


\section{Defining a Less Biased Sample of Companions}


\subsection{Host Sample Selection Effects}

High precision Doppler surveys are monitoring Sun-like stars for planetary companions and are necessarily sensitive enough 
to detect brown dwarfs and stellar companions within the same range of orbital period. However, to compare the relative 
abundances of stellar, brown dwarf and planetary companions, we cannot select our potential hosts from a non-overlapping 
union of the FGK spectral type target stars of the longest running, high precision Doppler surveys that are being monitored 
for planets \citep{Lineweaver03}. This is because Doppler survey target selection criteria often exclude close binaries 
(separation $< 2"$) from the target lists, and are not focused on detecting stellar companions. Some stars have also been 
left off the target lists because of high stellar chromospheric activity \citep{Fischer99}. These surveys are biased 
against finding stellar mass companions. We correct for this bias by identifying the excluded targets and then including 
in our sample any stellar companions from other Doppler searches found in the literature. Our sample selection is 
illustrated in Fig. \ref{fig:HR_25} and detailed in Table \ref{table:25pc} (complete list in the electronic version only) 
for stars closer than 25 pc and Fig. \ref{fig:HR_50} for stars closer than 50 pc.

\begin{figure}[!t]
\epsscale{0.75}
\plotone{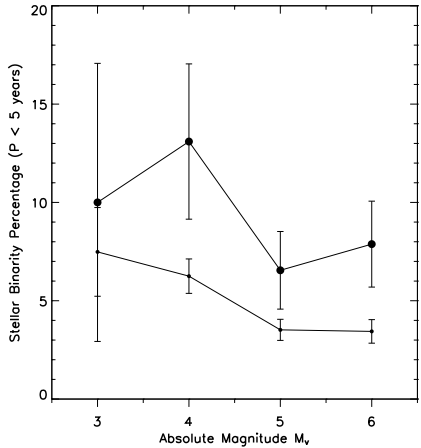}
\caption{Fraction of stars that are known to be close ($P < 5$ years) Doppler binaries as a function of absolute magnitude. 
For the 25 pc Sun-like sample (large dots), $\sim 11\%$ of stars are binaries and within the error bars, brighter stars do 
not appear to be significantly over-represented. If we include the extra stars to make the 50 pc Sun-like sample 
(small dots), the stellar binary fraction is lower and decreases as the systems get fainter.
}
\label{fig:LBV_Mv}
\end{figure} 

\begin{figure}[!t]
\epsscale{1.00}
\plotone{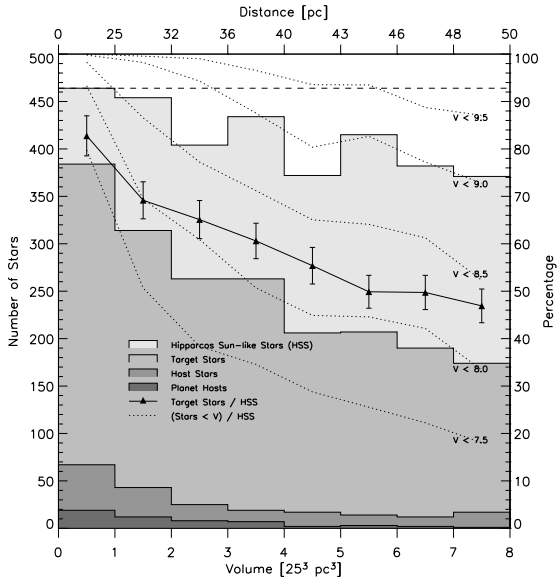}
\caption{Distance Dependence of Sample and Companions. Here we show the number of nearby Sun-like stars as a function of 
distance. Each histogram bin represents the stars in an equal volume spherical shell. Hence, a sample that is complete in 
distance out to 50 pc would produce a flat histogram (indicated by the horizontal dashed line). The lightest shade of grey 
represents Hipparcos Sun-like Stars out to 50 pc that fall within the parallelogram of Fig. \ref{fig:HR_50} (``HSS"). The 
next darker shade of grey represents Hipparcos stars that are being monitored for planets using the high 
precision Doppler techniques (8 groups described in Lineweaver \& Grether 2003). The triangles represent this number as a 
fraction of Hipparcos stars. This fraction needs to be as large as possible to minimize distance dependent selection 
effects in the target sample potentially associated with companion frequency. Also shown (darker grey) are the number of 
Hipparcos stars that have one or more companions in the mass range $10^{-3} < M/M_\odot < 1$, and those that host planets 
(darkest grey). Only those companions in the less-biased sample, $P < 5$ years and $M_2 > 10^{-3} M_\odot$ are shown 
(Section 2.2). The fraction of stars having an apparent magnitude $V$ brighter (lower) than a given value are shown by the 
5 dotted lines for $V < 7.5$ to $V < 9.5$.
}
\label{fig:LBV_volume}
\end{figure}

Most Doppler survey target stars come from the Hipparcos catalogue because host stars need to be both bright and have 
accurate masses for the Doppler method to be useful in determining the companion's mass. One could imagine that the Hipparcos 
catalogue would be biased in favor of binarity since hosts with bright close-orbiting stellar companions would be 
over-represented. We have checked for this over-representation by looking at the absolute magnitude dependence of the 
frequency of stellar binarity for systems closer than 25 and 50 pc (Fig. \ref{fig:LBV_Mv}). We found no significant 
decrease in the fraction of binaries in the dimmer stellar systems for the 25 pc sample and only a small decrease in the 
50 pc sample. Thus, the Hipparcos catalogue provides a good sample of potential hosts for our analysis, since it (i) 
contains the Doppler target lists as subsets (ii) is volume-limited for Sun-like stars out to $\sim 25$ pc \citep{Reid02} 
and (iii) it allows us to identify and correct for stars and stellar systems that were excluded. We limit our selection to 
Sun-like stars ($0.5 \leq B - V \leq 1.0$) or approximately those with a spectral type between F7 and K3. Following Udry 
(private communication) and the construction of the Coralie target list, we limit our anaylsis to main sequence stars, or 
those between -0.5 and +2.0 dex (below and above) an average main sequence value as defined by 
$5.4 (B - V) + 2.0 \leq M_v \leq 5.4 (B - V) - 0.5$. This sampled region, which we will call our ``Sun-like'' region of 
the HR diagram, is shown by the grey parallelograms in Figs. \ref{fig:HR_25} \& \ref{fig:HR_50}.

The Hipparcos sample is essentially complete to an absolute visual magnitude of $M_v = 8.5$ \citep{Reid02} within 25 pc of 
the Sun. Thus the stars in our 25 pc Sun-like sample represent a complete, volume-limited sample. In our sample we make 
corrections in companion frequency for stars that are not being targeted by Doppler surveys as well as corrections for 
mass and period companion detection selection effects (see Section 2.2). The result of these corrections is our 
less-biased distribution of companions to Sun-like stars within 25 pc. We also analyse a much larger sample of stars out to 
50 pc to understand the effect of distance on target selection and companion detection. Although less complete, with 
respect to the relative number of companions of different masses, the results from the 50 pc sample are similar to the 
results from the 25 pc sample (Section 3).


\begin{deluxetable*}{rccrcc}
\tablecolumns{6}
\tablewidth{300pt}
\tablecaption{Sun-like 25 pc Sample \label{table:25pc}}
\tablehead{
\colhead{Hipparcos} & \colhead{$B-V$} & \colhead{$M_V$} & \colhead{Distance} &
\colhead{Exoplanet} & \colhead{Companion} \\
\colhead{Number} & \colhead{} & \colhead{} & \colhead{(pc)} &
\colhead{Target} & \colhead{($P < 5$ years)} \\
\colhead{} & \colhead{} & \colhead{} & \colhead{} &
\colhead{} & \colhead{($M > M_{Jup}$)} }
\startdata
HIP 171      & 0.69         & 5.33         & 12.40        & Yes & \phn \\
HIP 518      & 0.69         & 4.44         & 20.28        & No & Star \\
HIP 544      & 0.75         & 5.39         & 13.70        & Yes & \phn \\
HIP 1031     & 0.78         & 5.68         & 20.33        & Yes & \phn \\
HIP 1292     & 0.75         & 5.36         & 17.62        & Yes & Planet \\
\enddata
\tablecomments{Table \ref{table:25pc} is published in its entirety in the electronic edition of the Astrophysical Journal. 
A portion is shown here for guidance regarding its form and content.}
\end{deluxetable*}


Stars in our Sun-like region are plotted as a function of distance in Fig. \ref{fig:LBV_volume}. Each histogram bin 
represents an equal volume spherical shell hence a sample complete in distance would produce a flat histogram. Also shown 
are the target stars, which are the subset of Hipparcos stars that are being monitored for planets by one of the 8 high 
precision Doppler surveys \citep{Lineweaver03} analysed here. The triangles in Fig. \ref{fig:LBV_volume} represent 
this number as a fraction of Hipparcos stars.

Since nearly all of the high precision Doppler surveys have apparent magnitude limited target lists (often $V < 7.5$), we 
investigate the effect this has on the total target fraction as a function of distance. The fraction of stars having an 
apparent magnitude $V$ brighter (lower) than a given value are shown by the 5 dotted lines for $V < 7.5$ to $V < 9.5$. 
For a survey, magnitude limited to $V = 7.5$, $80\%$ of the Sun-like Hipparcos stars will be observable between 0 pc and 
25 pc. This rapidly drops to only $20\%$ for stars between 48 and 50 pc. Thus the major reason why the target fraction 
drops with increasing distance is that the stars become too faint for the high precision Doppler surveys to monitor. The 
fact that the target fraction (triangles) lie near the $V < 8.0$ line indicates that on average $V \sim 8.0$ is the 
effective limiting magnitude of the targets monitored by the 8 combined high precision Doppler surveys.

In Fig. \ref{fig:HR_25}, $80 (=464-384)$ or $17\%$ of Hipparcos Sun-like stellar systems are not present in any of the 
Doppler target lists. The triangles in Fig. \ref{fig:HR_25} indicate that the ones left out are spread more or less evenly 
in B-V space spanned by the grey parallelogram. Similarly in Fig. \ref{fig:HR_50}, $1295 (=3296-2001)$ or $39\%$ are not 
included in any Doppler target list, but the triangles show that more K stars compared to FG stars have not been selected, 
again pointing out that the lower K dwarf stellar brightness is the dominant reason for the lower target fraction, not an 
effect strongly biased with respect to one set of companions over another.

In the Sun-like region of Fig. \ref{fig:HR_25} we use the target number (384) as the mother population for planets and 
brown dwarfs and the Hipparcos number (464) as the mother population for stars. To achieve the same normalizations for 
planetary, brown dwarf and stellar companions we assume that the fraction of these 384 targets that have exoplanet or 
brown dwarf companions is representative of the fraction of the 464 Hipparcos stars that have exoplanet or brown dwarf 
companions. Thus we renormalize the planetary and brown dwarf companions which have the target sample as their mother 
population to the Hipparcos sample by $464/384 = 1.21$ (``renormalization"). Since close-orbitting stellar companions are 
anti-correlated with close-orbitting sub-stellar companions and the 384 have been selected to exclude separations of $<2"$,
the results from the sample of 384 may be a slight over-estimate of the relative frequency of sub-stellar companions. 
However, this over-estimate will be less than $\sim 11\%$ because this is the frequency of close-orbitting stellar 
secondaries.

A non-overlapping sample of the 8 high precision Doppler surveys \citep{Lineweaver03} is used as the exoplanet target list 
where the Elodie target list was kindly provided by C. Perrier (private communication) and additional information to 
construct the Coralie target list from the Hipparcos catalogue was obtained from S. Udry (private communication). The Keck 
and Lick target lists are those of \citet{Nidever02}, since $\sim 7\%$ of the targets in \citet{Wright04} have not been 
observed over the full 5 year baseline used in this analysis. For more details about the sample sizes, observational 
durations, selection criteria and sensitivities of the 8 surveys see Table 4 of \citet{Lineweaver03}. 


\subsection{Companion Detection and Selection Effects}

The companions to the above Sun-like sample of host stars have primarily been detected using the Doppler technique (but not 
exclusively high precision exoplanet Doppler surveys) with some of the stellar pairs also being detected as astrometric 
or visual binaries. Thus we need to consider the selection effects of the Doppler method in order to define a less-biased 
sample of companions \citep{Lineweaver03}. As a consequence of the exoplanet surveys' limited monitoring duration we only 
select those companions with an orbital period $P < 5$ years. To reduce the selection effect due to the Doppler sensitivity 
we also limit our less-biased sample to companions of mass $M_2 > 0.001 M_\odot$. 

\begin{figure}[!t]
\plotone{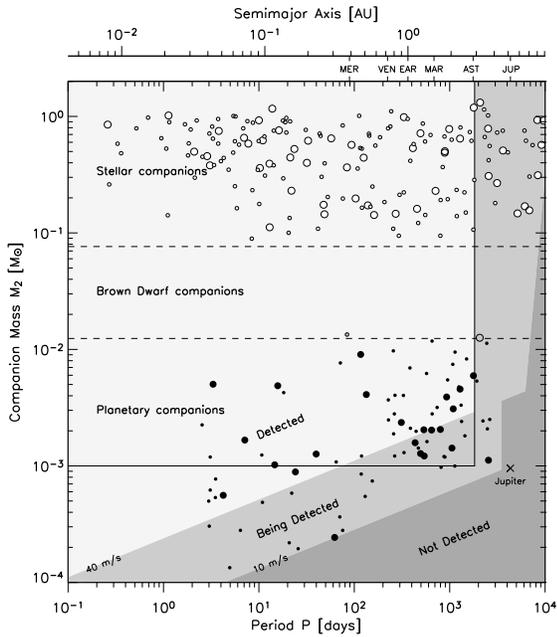}
\caption{Brown Dwarf Desert in Mass and Period. Estimated companion mass $M_2$ versus orbital period for the companions to 
Sun-like stars of our two samples: companions with hosts closer than 25 pc (large symbols) and those with hosts closer than 
50 pc, excluding those closer than 25 pc (small symbols). The companions in the thick solid rectangle are defined by 
periods $P < 5$ years, and masses $10^{-3} < M_2 \lsim M_\odot$, and form our less-biased sample of companions. The stellar 
(open circles), brown dwarf (grey circles) and planetary (filled circles) companions are separated by dashed lines at the 
hydrogen and deuterium burning onset masses of $80 \; M_{Jup}$ and $13 \;M_{Jup}$ respectively. This plot clearly shows the 
brown dwarf desert for the $P < 5$ year companions. Planets are more frequent at larger periods than at shorter periods 
(see Fig. \ref{fig:Period_25_50}). The ``Detected", ``Being Detected" and ``Not Detected" regions of the mass-period space 
show the extent to which the high precision Doppler method is currently able to find companions \citep{Lineweaver03}. 
see Appendix for discussion of $M_2$ mass estimates.
}
\label{fig:Binary_m_P}
\end{figure}

\begin{figure}[!t]
\plotone{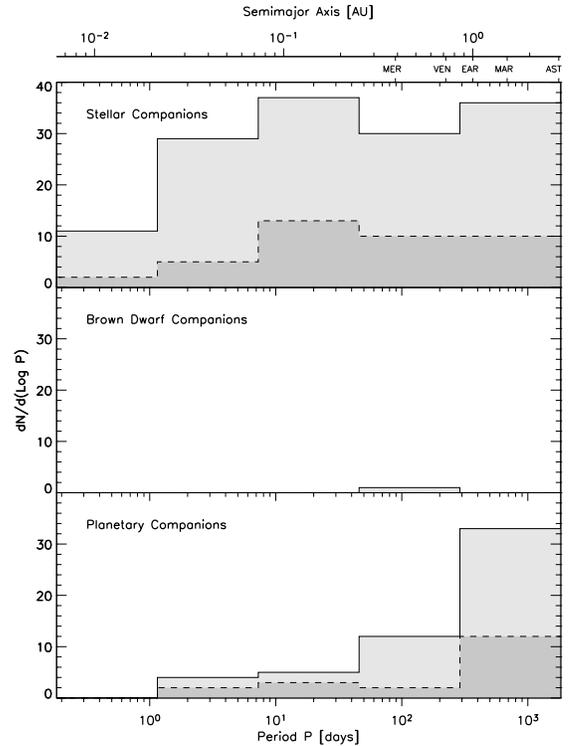}
\caption{Projection of Fig. \ref{fig:Binary_m_P} onto the period axis for the 25 pc (dark grey) and 50 pc (light grey) 
samples. Planets are more clumped towards higher periods than are stellar companions. This would be a selection effect with 
no significance if the efficiency of finding short period stellar companions with the low precision Doppler technique used 
to find spectroscopic binaries, was much higher than the efficiency of finding exoplanets with high precision spectroscopy. 
\citet{Konacki04} and \citet{Pont04} conclude that the fact that the transit photometry method has found planets in sub 2.5 
day periods (while the Doppler method has found none) is due to higher efficiency for small periods and many more target 
stars and thus that these two observations do not conflict. Thus there seems to be a real difference in the period 
distributions of stellar and planetary companions.
}
\label{fig:Period_25_50}
\end{figure}

Fig. \ref{fig:Binary_m_P} shows all of the Doppler companions to the Sun-like 25 pc and 50 pc samples within the mass and 
period range considered here. Our less-biased companions are enclosed by the thick solid rectangle. Given a fixed number of 
targets, the ``Detected" region should contain all companions that will be found for this region of mass-period space. The 
``Being Detected" region should contain some but not all companions that will be found in this region and the 
``Not Detected" region contains no companions since the current Doppler surveys are either not sensitive enough or have 
not been observing for a long enough duration. To avoid the incomplete ``Being Detected" region we limit our sample of 
companions to $M_2 > 0.001 M_\odot$. In \citet{LGH03} we describe a crude method for making a completeness correction for 
the lower right corner of the solid rectangle falling within the ``Being Detected'' region. The result for the $d < 25$ pc 
sample is a one planet correction to the lowest mass bin and for the $d < 50$ pc sample, a six planet correction to the 
lowest mass bin (see Table \ref{table:companions} - footnote b). Fig. \ref{fig:Period_25_50} shows a projection of 
Fig. \ref{fig:Binary_m_P} onto the period axis. Planets are more clumped towards higher periods than are stellar companions. 
The Doppler planet detection method is not biased against short period planets. The Doppler stellar companion detections 
are not significantly biased for shorter periods or against longer periods in our samples analysis range (period $< 5$ years) 
since Doppler instruments of much lower precision than those used to detect exoplanets are able to detect any Doppler 
companions of stellar mass. Thus this represents a real difference in period distributions between stellar and planetary 
companions.

The companions in Fig. \ref{fig:Binary_m_P} all have radial velocity (Doppler) solutions. Some of the companions also have 
additional photometric, interferometric, astrometric or visual solutions. The exoplanet Doppler orbits are taken from the 
Extrasolar Planets Catalog \citep{EPC}. Only the planet orbiting the star HIP 108859 (HD 209458) has an additional 
photometric solution but this companion falls outside our less-biased region ($M_2 < M_{Jup}$). For the stellar companion 
data, the single-lined (SB1) and double-lined (SB2) spectroscopic binary orbits are primarily from the Ninth Catalogue of 
Spectroscopic Binary Orbits \citep{SB9} with additional interferometric, astrometric or visual solutions from the 
6th Catalog of Orbits of Visual Binary Stars (Washington Double Star Catalog, Hartkopf \& Mason 2004). Many additional 
SB1s come from \citet{Halbwachs03}. Stellar binaries and orbital solutions also come from 
\citet{Endl04, Halbwachs00, Mazeh03, Tinney01, Jones02, Vogt02, Zucker01a}.

We examine the inclination distribution for the 30 Doppler companions ($d < 50$ pc) with an astrometric or visual 
solution. We find that 24 of these 30 companions have a minimum mass larger than $80 M_{Jup}$ (Doppler stellar candidates) 
and that 6 of these 30 companions have a minimum mass between $13 M_{Jup}$ and $80 M_{Jup}$ (Doppler brown dwarf candidates). 
These 6 Doppler brown dwarf candidates are a subset of the 16 Doppler brown dwarf candidates in the far sample that have 
an astrometric orbit derived with a confidence level greater than $95\%$ from Hipparcos measurements 
\citep{Halbwachs00, Zucker01a} and are thus assumed to have an astrometric orbit. 

\begin{figure}[!t]
\plotone{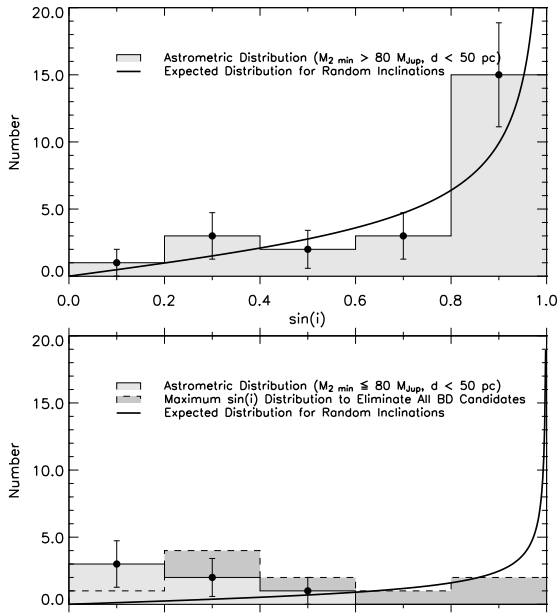}
\caption{
Astrometric inclination distribution for close companions ($d < 50$ pc) with a minimum mass larger than $80 M_{Jup}$ 
(Doppler stellar candidates - TOP) and between $13 M_{Jup}$ and $80 M_{Jup}$ (Doppler brown dwarf candidates - BOTTOM). 
There are 24 companions with astrometric solutions and a minimum mass in the stellar regime and 6 with a minimum mass 
in the brown dwarf regime. The inclination distribution is approximately random for companions with a minimum mass in the 
stellar regime whereas it is biased towards low inclinations for companions in the brown dwarf regime. All 6 astrometric 
determinations of $sin(i)$ for brown dwarf candidates put their true mass in the stellar regime. Also shown is the 
distribution of the maximum values of $sin(i)$ that would place the true masses of the remaining 10 brown dwarf candidates 
without astrometric or visual solutions in the stellar regime. A distribution less-biased than the observed $sin(i)$ 
distribution would be required. This strongly suggests that the 10 candidates without astrometric or visual solutions 
will also have masses in the stellar regime. Therefore, astrometric corrections leave us with no solid candidates 
with masses in the brown dwarf region. Two weak brown dwarf candidates are worth mentioning. HD 114762 has a minimum mass 
below $13 M_{Jup}$. However, to convert minimum mass to mass, we have assumed random inclinations and have used 
$<sin(i)> \approx 0.785$. This conversion puts the estimated mass of HD 114762 in the brown dwarf regime 
($M \gsim \: 13 M_{Jup}$). In Fig.\ \ref{fig:Binary_m_P}, this is the only companion lying in the brown dwarf regime. 
Another weak brown dwarf candidate is the only candidate that requires a $sin(i) < 0.2$ to place its mass in the stellar 
regime.
}
\label{fig:N_sini}
\end{figure}

As shown in Fig. \ref{fig:N_sini}, the inclination distribution is approximately random for the 24 companions with a 
minimum mass in the stellar regime whereas it is biased towards low inclinations for the 6 companions in the brown dwarf 
regime. All 6 of the Doppler brown dwarf candidates with an astrometric determination of their inclination have a true 
mass in the stellar regime. This includes all 3 of the Doppler brown dwarf candidates that are companions to stars in our 
close sample ($d < 25$ pc) thus leaving an empty brown dwarf regime. Also shown in Fig. \ref{fig:N_sini}, is the 
distribution of the maximum values of $sin(i)$ that would put the true masses of the remaining 10 Doppler brown dwarf 
candidates with unknown inclinations in the stellar regime. This distribution is substantially less-biased than the 
observed $sin(i)$ distribution, strongly suggesting that the remaining 10 Doppler brown dwarf candidates will also have 
masses in the stellar regime. Thus astrometric corrections leave us with no solid candidates with masses in the brown 
dwarf regime from the 16 Doppler brown dwarf candidates in the far sample ($d < 50$ pc), consistent with the result 
obtained for the close sample.

\begin{table*}[!htp]
\begin{center}
\caption{Hipparcos Sample, Doppler Targets and Detected Companions for Near and Far Samples}
\label{table:companions}
\scriptsize
\begin{tabular}{|c|c|c|c|c|c|c|c|c|c|}
\hline
Sample & Hipparcos & \multicolumn{2}{c}{Doppler} \vline & \multicolumn{6}{c}{Companions} \vline \\ 
\cline{5-10}
       & Number    & \multicolumn{2}{c}{Target} \vline  & Total & Planets & BDs & \multicolumn{3}{c}{Stars} \vline \\
\cline{3-4} \cline{8-10}
       &           & Number  & \%$^{a}$                 &       &         &     & Total & SB1 & SB2 \\
\hline
$d < 25$ pc & 1509 & 627 & 42\% & - & 22 & - & - & - & - \\
Sun-like    & 464  & 384 & 83\% & 59 (+15)$^{f}$ & 19 (+1$^{b}$,+4$^{c}$) & 0 & 40 (+7$^{d}$,+3$^{e}$) & 
25 (9)$^{h}$ & 15 (8)$^{h}$ \\
Dec $< 0 \degr$    & 211 & 211 & 100\% & 20 (+10)$^{f}$ & 10 & 0 & 10 (+7$^{d}$,+3$^{e}$) & 
8 (3)$^{h}$ & 2 (1)$^{h}$ \\
Dec $\geq 0 \degr$ & 253 & 173 & 68\% & 39 & 9 & 0 & 30 & 
17 (6)$^{h}$ & 13 (7)$^{h}$ \\
\hline
$d < 50$ pc & 6924 & 2351 & 34\% & - & 58 & - & - & - & - \\
Sun-like    & 3296 & 2001 & 61\% & 198 (+80)$^{f}$ & 54 (+6$^{b}$,+19$^{e}$) & 1$^{g}$ & 143 (+14$^{d}$,+41$^{e}$) & 
90 (18)$^{h}$ & 53 (12)$^{h}$ \\
Dec $< 0 \degr$    & 1647 & 1525 & 93\% & 72 (+74)$^{f}$ & 33 (+19)$^{e}$ & 0 & 39 (+14$^{d}$,+41$^{e}$) & 
27 (7)$^{h}$ & 12 (2)$^{h}$ \\
Dec $\geq 0 \degr$ & 1649 & 476  & 29\% & 126 & 21 & 1 & 104 & 
63 (11)$^{h}$ & 41 (10)$^{h}$ \\
\tableline 
\end{tabular}
\end{center}
\scriptsize
\noindent
$^{a}$ Percentage of Hipparcos stars that are Doppler targets. \\
$^{b}$ Completeness correction in the lowest mass bin for the lower right corner of our sample in Fig. 5 lying in the 
``Being Detected" region (see Lineweaver, Grether \& Hidas 2003). \\
$^{c}$ Renormalization for planetary target population (384) being less than stellar companion mother population (464) 
(see discussion in Section 2.1). \\
$^{d}$ Correction based on the most likely scenario that the southern stellar companions from \citet{Jones02} have periods 
$< 5$ years. \\
$^{e}$ Correction for north/south declination asymmetry in companion fraction after correcting for \citet{Jones02} 
detections (see Section 2.2). \\
$^{f}$ Total of corrections b through to e. \\
$^{g}$ Result from assuming $<sin(i)> = 0.785$ when $i$ is unknown (see caption of Fig. 7 and Appendix). \\
$^{h}$ Number of these spectroscopic binaries with an additional astrometric or visual solution (see Appendix).\\
\end{table*}

The size of the 25 pc and 50 pc samples, the extent to which they are being targeted for planets, and the number and types 
of companions found along with any associated corrections are summarised in Table \ref{table:companions}. For the stars 
closer than 25 pc, 59 have companions in the less-biased region (rectangle circumscribed by thick line) of 
Fig. \ref{fig:Binary_m_P}. Of these, 19 are exoplanets, 0 are brown dwarfs and 40 are of stellar mass. Of the stellar 
companions, 25 are SB1s and 15 are SB2s. For the stars closer than 50 pc, 198 have companions in the less-biased region. 
Of these, 54 are exoplanets, 1 is a brown dwarf and 143 are stars. Of the stellar companions, 90 are SB1s and 53 are SB2s.

We find an asymmetry in the north/south declination distribution of the Sun-like stars with companions, probably
due to undetected or unpublished stellar companions in the south. The number of hosts closer than 25 pc with planetary or 
brown dwarf companions are symmetric in north/south declination to within one sigma Poisson error bars, but because more 
follow up work has been done in the north, more of the hosts with stellar companions with orbital solutions are in the 
northern hemisphere (30) compared with the southern (10). A comparison of our northern sample of hosts with stellar 
companions to the similarly selected approximately complete sample of \citet{Halbwachs03} indicates that our 25 pc 
northern sample of hosts with stellar companions is also approximately complete. Under this assumption, the number of 
stellar companions missing from the south can be estimated by making a minimal correction up to the one sigma error level 
below the expected number, based on the northern follow up results. Of the 464 Sun-like stars closer than 25 pc, 211 have 
a southern declination (Dec $< 0 \degr$) and 253 have a northern declination (Dec $\geq 0 \degr$) and thus 
$\sim 25 (25/211 \approx 30/253)$ stars in the south should have a stellar companion when fully corrected or 20 if we 
make a minimal correction. Thus we estimate that we are missing at least $\sim 10 (= 20-10)$ stellar companions in the 
south, 7 of which have been detected by \citet{Jones02} under the plausible assumption that the orbital periods of the 
companions detected by \citet{Jones02} are less than 5 years. Although these 7 SB1 stellar companions detected by 
\citet{Jones02} have as yet no published orbital solutions, we assume that the SB1 stellar companions detected by 
\citet{Jones02} have $P < 5$ years since they have been observed as part of the high Doppler precision program at the 
Anglo-Australian Observatory (started in 1998) for a duration of less than 5 years before being announced. The 
additional estimated stellar companions are assumed to have the same mass distribution as the other stellar companions. 

We can similarly correct the declination asymmetry in the sample of Sun-like stars closer than 50 pc. We find that there 
should be, after a minimal correction, an additional 55 stars that are stellar companion hosts in the southern hemipshere. 
14 of these 55 stellar companions are assumed to have been detected by \citet{Jones02}. An asymmetry found in the planetary 
companion fraction in the 50 pc sample due to the much larger number of stars being monitored less intensively for 
exoplanets in the south ($\sim 2\% = 33/1525$) compared to the north ($\sim 4\% = 21/476$) results in a correction of 
19 planetary companions in the south. The results given in Table \ref{table:minima} are done both with and without the 
asymmetry corrections.

Unlike the 25 pc sample for which we are confident that the small corrections made to the number of companions will 
result in a reliable estimate of a census, correcting the 50 pc sample for the large number of missing companions is 
less reliable. This is so because if it were complete, the 50 pc sample would have approximately 8 times the number of 
companions as the 25 pc sample, since the 50 pc sample has 8 times the volume of the 25 pc sample. However, the incomplete 
50 pc sample has only $\sim 7 (=3296/464)$ times the number of Hipparcos stars, $\sim 5 (=2001/384)$ times as many 
exoplanet targets and $\sim 3$ times as many companions as the 25 pc sample. Thus rather than correcting both planetary 
and stellar companions by large amounts we show in Section 3 that the relative number and distribution of the observed 
planetary and stellar companions (plus a small completeness correction for the ``Being Detected" region of 6 planets and 
an additional 14 probable stellar companions from \citet{Jones02} - see Table \ref{table:companions}) remains approximately 
unchanged when compared to the corrected companion distribution of the 25 pc sample. Analyses both with and without a 
correction for the north/south asymmetry produce similar results for the brown dwarf desert (Table \ref{table:minima}).


\section{Companion Mass Function}

\begin{figure}[!t]
\epsscale{1.0}
\plotone{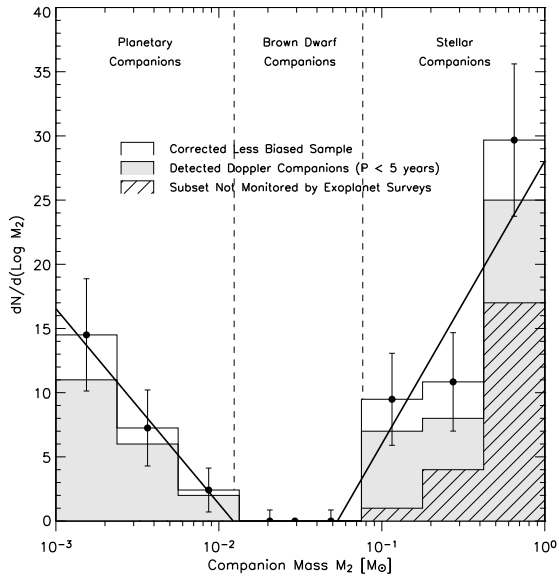}
\caption{Brown Dwarf Desert in Close Sample. Histogram of the companions to Sun-like stars closer than 25 pc plotted 
against mass. The grey histogram is made up of Doppler detected companions in our less-biased ($P < 5$ years and 
$M_2 > 10^{-3} M_\odot$) sample. The corrected version of this less-biased sample includes an extra 7 probable SB1 stars 
from \citep{Jones02} (Table \ref{table:companions} - footnote d) and an extra 3 stars from an asymmetry in the host 
declination distribution (Table \ref{table:companions} - footnote e). The planetary mass companions are also renormalized 
to account for the small number of Hipparcos Sun-like stars that are not being Doppler monitored (21\% renormalization, 
Table \ref{table:companions} - footnote c) and a 1 planet correction for the undersampling of the lowest mass bin due to 
the overlap with the ``Being Detected'' region (Table \ref{table:companions} - footnote b). The hatched histogram is the 
subset of detected companions to hosts that are not included on any of the exoplanet search target lists and hence shows 
the extent to which the exoplanet target lists are biased against the detection of stellar companions. Since instruments 
with a radial velocity sensitivity $K_S \leq 40$ m/s (see Eq. 2 of Appendix) were used for all the companions, we expect 
no other substantial biases to affect the relative amplitudes of the stellar companions on the right-hand side (RHS) and 
the planetary companions on the left-hand side (LHS). The brown dwarf mass range is empty.
}
\label{fig:mass_25}
\end{figure}

\begin{figure}[!t]
\plotone{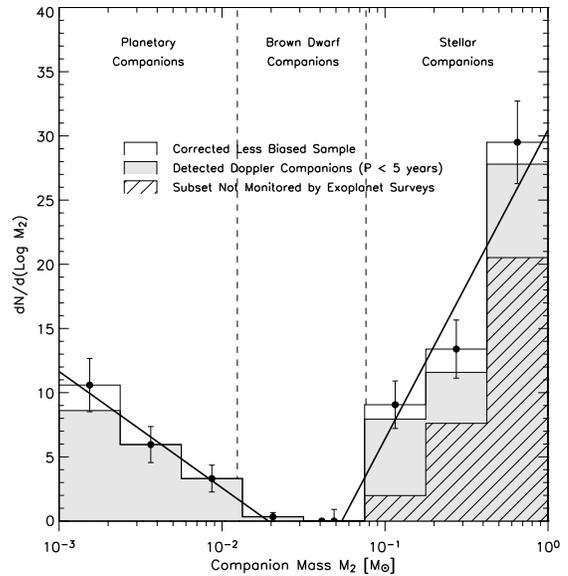}
\caption{Same as Fig. \ref{fig:mass_25} but for the larger 50 pc sample renormalized to the size of the 25 pc sample.
Fitting straight lines using a weighted least squares fit to the 3 bins on the LHS and RHS, gives us gradients of 
$-9.1 \pm 2.9$ and $24.1 \pm 4.7$ respectively (solid lines). Hence the brown dwarf desert is significant at more than the 
3 sigma level. These LHS and RHS slopes agree to within about 1 sigma of those in Fig. \ref{fig:mass_25}. The ratio of the 
number of companions on the LHS to the RHS is also about the same for both samples. Hence the relative number and 
distribution of companions is approximately the same as in Fig. \ref{fig:mass_25}. The separate straight line fits to the 
3 bins on the LHS and RHS intersect at $M = 43 \:^{+14}_{-23} M_{Jup}$ beneath the abscissa. Approximately 16\% of the 
stars have companions in our less-biased region. Of these, $4.3\% \pm 1.0\%$ have companions of planetary mass, 
$0.1 \:^{+0.2}_{-0.1}\%$ have brown dwarf companions and $11.2\% \pm 1.6\%$ have companions of stellar mass. We 
renormalize the mass distribution in this figure by comparing each bin in this figure with its corresponding bin in 
Fig. \ref{fig:mass_25} and scaling the vertical axis of Fig. \ref{fig:mass_50} so that the difference in height between 
the bins is on average a minimum. We find that the optimum renormalization factor is 0.33. This plot does not include the 
asymmetry correction for the planetary and stellar companions discussed in Section 2.2 and shown in 
Table \ref{table:companions}. 
}
\label{fig:mass_50}
\end{figure}

The close companion mass function to Sun-like stars clearly shows a brown dwarf desert for both the 25 pc 
(Fig. \ref{fig:mass_25}) and the 50 pc (Fig. \ref{fig:mass_50}) samples. The numbers of both the planetary and stellar mass 
companions decrease toward the brown dwarf mass range. Both plots contain the detected Doppler companions, shown as the 
grey histogram, within our less-biased sample of companions ($P < 5$ years and $M_2 > 10^{-3} M_\odot$, see Section 2.2). 
The hatched histograms at large mass show the subset of the stellar companions that are not included in any of the exoplanet 
Doppler surveys. A large bias against stellar companions would have been present if we had only included companions found 
by the exoplanet surveys. For multiple companion systems, we select the most massive companion in our less biased sample 
to represent the system. We put the few companions (3 in the 25 pc sample, 6 in the 50 pc sample) that have a mass 
slightly larger than $1 \: M_\odot$ in the largest mass bin in the companion mass distributions.

Fitting straight lines using a weighted least squares method to the 3 bins on the left-hand side (LHS) and right-hand side 
(RHS) of the brown dwarf region of the mass histograms (Figs. \ref{fig:mass_25} \& \ref{fig:mass_50}), gives us gradients 
of $-15.2 \pm 5.6$ (LHS) and $22.0 \pm 8.8$ (RHS) for the 25 pc sample and $-9.1 \pm 2.9$ (LHS) and $24.1 \pm 4.7$ (RHS) 
for the 50 pc sample. Since the slopes have opposite signs, they form a valley which is the brown dwarf desert. The 
presence of a valley between the negative and positive sloped lines is significant at more than the 3 sigma level. 
The ratio of the corrected number of companions in the less-biased sample on the LHS to the RHS along with their 
poisson error bars is $(24 \pm 9)/(50 \pm 13) = 0.48 \pm 0.22$ with no companions in the middle 2 bins for the 25 pc 
sample. For the larger 50 pc sample the corrected less-biased LHS/RHS ratio is $(60 \pm 14)/(157 \pm 22) = 0.38 \pm 0.10$, 
with 1 brown dwarf companion in the middle 2 bins. Thus the LHS and RHS slopes agree to within about 1 sigma and so do the 
LHS/RHS ratios, indicating that the companion mass distribution for the larger 50 pc sample is not significantly different 
from the more complete 25 pc sample and that the relative fraction of planetary, brown dwarf and stellar companions is 
approximately the same. A comparison of the relative number of companions in each bin in Fig. \ref{fig:mass_25} with its 
corresponding bin in Fig. \ref{fig:mass_50} produces a best-fit of ${\chi}^2 = 1.9$.

To find the driest part of the desert, we fit separate straight lines to the 3 bins on either side of the brown dwarf 
desert (solid lines) in Figs. \ref{fig:mass_25} \& \ref{fig:mass_50}. The deepest part of the valley where the straight 
lines cross beneath the abscissa is at $M = 31 \:^{+25}_{-18} M_{Jup}$ and $M = 43 \:^{+14}_{-23} M_{Jup}$ for the 
25 and 50 pc samples respectively. These results are summarized in Table \ref{table:minima}. The driest part of the desert 
is virtually the same for both samples even though we see a bias in the stellar binarity fraction of the 50 pc sample 
(Fig. \ref{fig:LBV_Mv}). We have done the analysis with and without the minimal declination asymmetry correction. 
The position of the brown dwarf minimum and the slopes are robust to this correction (see Table \ref{table:minima}).

The smaller 25 pc Sun-like sample contains 464 stars with $16.0\% \pm 5.2\%$ of these having companions in our corrected 
less-biased sample. Of these $\sim 16\%$ with companions, $5.2\% \pm 1.9\%$ are of planetary mass and $10.8\% \pm 2.9\%$ 
are of stellar mass. None is of brown dwarf mass. This agrees with previous estimates of stellar binarity such as that 
found by \citet{Halbwachs03} of $14\%$ for a sample of G-dwarf companions with a slightly larger period range 
($P < 10$ years). The planet fraction agrees with the fraction $4\% \pm 1\%$ found in \citet{Lineweaver03} when most of 
the known exoplanets are considered. The 50 pc sample has a large incompleteness due to the lower fraction of monitored 
stars (Fig. \ref{fig:LBV_volume}) but as shown above, the relative number of companion planets, brown dwarfs and stars is 
approximately the same as for the 25 pc sample. The 50 pc sample has a total companion fraction of 
$15.6\% \pm 2.8\%$, where $4.3\% \pm 1.0\%$ of the companions are of planetary mass, $0.1^{+0.2}_{-0.1}\%$ are of brown 
dwarf mass and $11.2\% \pm 1.6\%$ are of stellar mass. Table \ref{table:fraction} summarizes these companion fractions.

Surveys of the multiplicity of nearby Sun-like stars yield the relative numbers of single, double and multiple star 
systems. According to \citet{Duquennoy91}, $51\%$ of star systems are single stars, $40\%$ are double star systems, $7\%$ 
are triple and $2\%$ are quadruple or more. Of the $49\% (= 40 + 7 + 2)$ which are stellar binaries or multiple star 
systems, $11\%$ have stellar companions with periods less than 5 years and thus we can infer that the remaining $38\%$ have 
stellar companions with $P > 5$ years. Among the $51\%$ without stellar companions, we find that $\sim 5\%$ have close 
($P < 5$ years) planetary companions with $1 < M/M_{Jup} < 13$, while $<1\%$ have close brown dwarfs companions.

\begin{figure}[!t]
\plotone{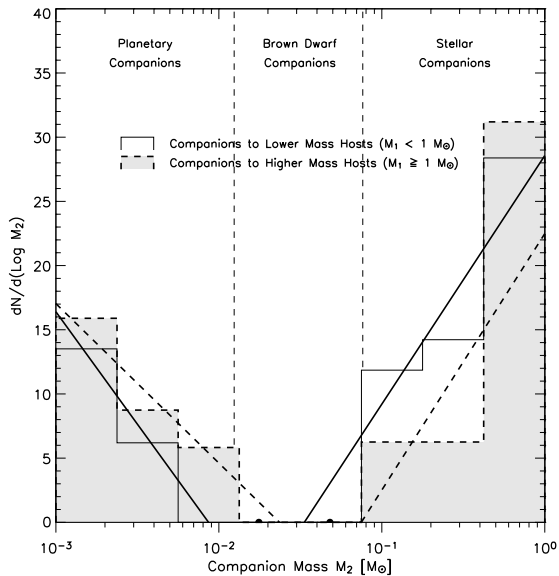}
\caption{Same as Fig. \ref{fig:mass_25} but for the 25 pc sample split into companions to lower mass hosts ($M_1 < 1 
M_\odot$) and companions to higher mass hosts ($M_1 \geq 1 M_\odot$). The lower mass hosts have $4.2\%$ planetary, $0.0\%$ 
brown dwarf and $11.8\%$ stellar companions. The higher mass hosts have $6.6\%$ planetary, $0.0\%$ brown dwarf and $9.4\%$ 
stellar companions. The Doppler method should preferentially find planets around lower mass stars where a greater radial 
velocity is induced. This is the opposite of what we observe. To aid comparison, both samples are scaled such that they 
contain the same number of companions as the full corrected less-biased 25 pc sample of Fig. \ref{fig:mass_25}. 
}
\label{fig:mass_host}
\end{figure}

\begin{figure}[!ht]
\plotone{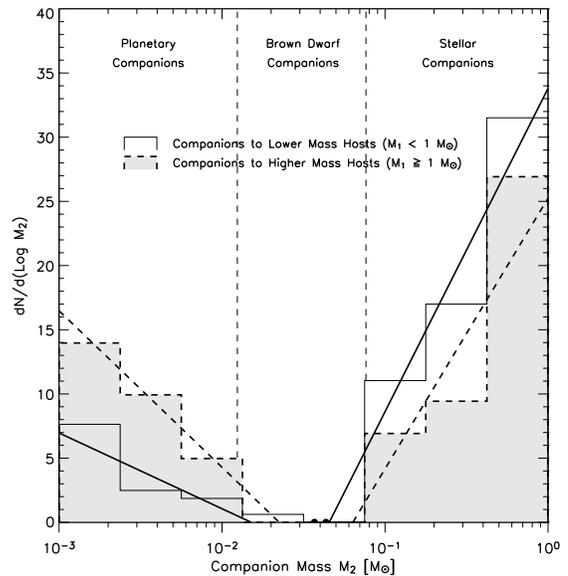}
\caption{Same as Fig. \ref{fig:mass_50} but for the 50 pc sample split into companions to lower mass hosts ($M_1 < 1 
M_\odot$) and companions to higher mass hosts ($M_1 \geq 1 M_\odot$). Both samples are scaled such that they contain the 
same number of companions as the corrected less-biased 50 pc sample of Fig. \ref{fig:mass_50}. Also shown are the linear
best-fits to the planetary and stellar companions of the two populations.
}
\label{fig:mass_host_50}
\end{figure}

The Doppler method should preferentially find planets around lower mass stars where a greater radial velocity is induced. 
This is the opposite of what is observed as shown in Figs. \ref{fig:mass_host} and \ref{fig:mass_host_50} where we split 
the 25 and 50 pc samples respectively into companions to hosts with masses above and below $1 \: M_\odot$. We scale these 
smaller samples to the size of the full 25 and 50 pc samples (Figs. \ref{fig:mass_25} and \ref{fig:mass_50} respectively). 
The Doppler technique is also a function of $B-V$ color \citep{Saar98} with the level of systematic errors in the radial 
velocity measurements, decreasing as we move from high mass to low mass ($B-V = 0.5$ to $B-V = 1.0$) through our two 
samples, peaking for late K spectral type stars before increasing for the lowest mass M type stars again. Hence again 
finding planets around the lower mass stars (early K spectral type) in our sample should be easier.


\begin{table*}[!htp]
\begin{center}
\caption{Companion Slopes and Companion Desert Mass Minima}
\label{table:minima}
\scriptsize
\begin{tabular}{|l|c|c|c|c|c|}
\tableline
Sample & Asymmetry & Figure & LHS slope & RHS slope & Slope Minima$^{a}$ \\
       & Correction &      &      &     & [$M_{Jup}$] \\
\tableline
$d < 25$ pc & Yes & 8 & $-15.2 \pm 5.6$ & $22.0 \pm 8.8$ & $31 \:^{+25}_{-18}$ \\
$d < 25$ pc & No & & $-15.2 \pm 5.6$ & $20.7 \pm 8.5$ & $30 \:^{+25}_{-17}$ \\
\tableline
$d < 50$ pc & Yes & & $-9.4 \pm 3.0$ & $24.3 \pm 4.6$ & $44 \:^{+15}_{-24}$ \\
$d < 50$ pc & No & 9 & $-9.1 \pm 2.9$ & $24.1 \pm 4.7$ & $43 \:^{+14}_{-23}$ \\
\tableline
$d < 25$ pc \& $M_1 < 1 M_\odot$ & Yes & 10 & $-17.5 \pm 5.4$ & $19.4 \pm 10.7$ & $18 \:^{+17}_{-9}$ \\
$d < 50$ pc \& $M_1 < 1 M_\odot$ & No & 11 & $-5.9 \pm 5.1$ & $25.2 \pm 11.4$ & $39 \:^{+9}_{-23}$ \\
\tableline
$d < 25$ pc \& $M_1 \geq 1 M_\odot$ & Yes & 10 & $-12.4 \pm 9.2$ & $20.0 \pm 10.9$ & $50 \:^{+28}_{-26}$ \\
$d < 50$ pc \& $M_1 \geq 1 M_\odot$ & No & 11 & $-12.2 \pm 8.2$ & $21.1 \pm 10.4$ & $45 \:^{+21}_{-21}$ \\
\tableline
\end{tabular} 
\end{center}
\noindent
\scriptsize
$^{a}$ values of mass where the best-fitting lines, to the LHS and RHS, intersect. The errors given are from the range 
between the two intersections with the abscissa.\\
\end{table*}


\begin{table*}[!htp]
\begin{center}
\caption{Companion Fraction Comparison}
\label{table:fraction}
\scriptsize
\begin{tabular}{|l|c|c|c|c|c|c|}
\hline
Sample & Asymmetry & Figure & Total $\%$ & Planetary $\%$ & Brown Dwarf $\%$ & Stellar $\%$ \\
       & Correction &       &            &                &                  &              \\
\tableline
$d < 25$ pc & Yes & 8 & $16.0 \pm 5.2$ & $5.2 \pm 1.9$ & $0.0 \:^{+0.4}_{-0.0}$ & $10.8 \pm 2.9$ \\
$d < 25$ pc & No  &   & $15.3 \pm 5.0$ & $5.2 \pm 1.9$ & $0.0 \:^{+0.4}_{-0.0}$ & $10.1 \pm 2.7$ \\
\tableline
$d < 50$ pc & Yes &   & $15.6 \pm 2.8$ & $4.4 \pm 1.0$ & $0.1 \:^{+0.2}_{-0.1}$ & $11.1 \pm 1.6$ \\
$d < 50$ pc & No  & 9 & $15.6 \pm 2.8$ & $4.3 \pm 1.0$ & $0.1 \:^{+0.2}_{-0.1}$ & $11.2 \pm 1.6$ \\
\tableline
$d < 25$ pc \& $M_1 < 1 M_\odot$ & Yes & 10 & $16.0 \pm 5.8$ & $4.2 \pm 1.9$ & $0.0 \:^{+0.4}_{-0.0}$ & $11.8 \pm 3.5$ \\
$d < 50$ pc \& $M_1 < 1 M_\odot$ & No  &11 & $15.6 \pm 6.0$ & $2.6 \pm 1.7$ & $0.2 \:^{+0.4}_{-0.2}$ & $12.8 \pm 3.9$ \\
\tableline
$d < 25$ pc \& $M_1 \geq 1 M_\odot$ & Yes & 10 & $16.0 \pm 7.0$ & $6.6 \pm 3.1$ & $0.0 \:^{+0.4}_{-0.0}$ & $9.4 \pm 3.5$ \\
$d < 50$ pc \& $M_1 \geq 1 M_\odot$ & No  & 11 & $15.6 \pm 6.7$ & $6.2 \pm 2.9$ & $0.0 \:^{+0.4}_{-0.0}$ & $9.4 \pm 3.4$ \\
\tableline 
\end{tabular}
\end{center}
\end{table*}

\subsection{Comparison with Other Results}

Although there are some similarities, the companion mass function found by \citet{Heacox99, Zucker01b, Mazeh03} is 
different from that shown in Figs. \ref{fig:mass_25} \& \ref{fig:mass_50}. Our approach was to normalize the companion 
numbers to a well-defined sub-sample of Hipparcos stars whereas these authors use two different samples of stars, one to 
find the planetary companion mass function and another to find the stellar companion mass function, which are then 
normalized to each other. The different host star properties and levels of completeness of the two samples may make this 
method more prone than our method, to biases in the frequencies of companions.

Both \citet{Heacox99} and \citet{Zucker01b} combined the companions of the stellar mass sample of \citet{Duquennoy91} with 
the known substellar companions, but identified different mass functions for the planetary mass regime below 10 $M_{Jup}$ 
and similar flat distributions in logarithmic mass for brown dwarf and stellar mass companions. \citet{Heacox99} found that 
the logarithmic mass function in the planetary regime is best fit by a power-law ($dN/dlogM \propto M^{\Gamma}$) with index 
$\Gamma$ between 0 and -1 whereas \citet{Zucker01b} find an approximately flat distribution (power-law with index 0). Our 
work here and in \citet{Lineweaver03} suggests that neither the stellar nor the planetary companion distributions are flat 
($\Gamma = -0.7$). Rather, they both slope down towards the brown dwarf desert, more in agreement with the results of 
\citet{Heacox99}.

The work most similar to ours is probably \citep{Mazeh03} who looked at a sample of main sequence stars with primaries in 
the range $0.6 - 0.85 \: M_\odot$ and $P < 3000$ days using infrared spectroscopy and combined them with the known 
substellar companions of these main sequence stars and found that in logarithmic mass the stellar companions reduce in 
number towards the brown dwarf mass range. This agrees with our results for the shape of the stellar mass companion 
function. However, they identify a flat distribution for the planetary mass companions in contrast to our non-zero slope 
(see Table \ref{table:minima}). \citet{Mazeh03} found the frequency of stellar and planetary companions 
($M_2 > 1 \: M_{Jup}$) to be $15\%$ (for stars below $0.7 \: M_\odot$) and $3\%$ respectively. This compares with our 
estimates of $8\%$ (for stars below $0.7 \: M_\odot$) and $5\%$. The larger period range used by \citet{Mazeh03} can 
account for the difference in stellar companion fractions.


\begin{figure}[!t]
\plotone{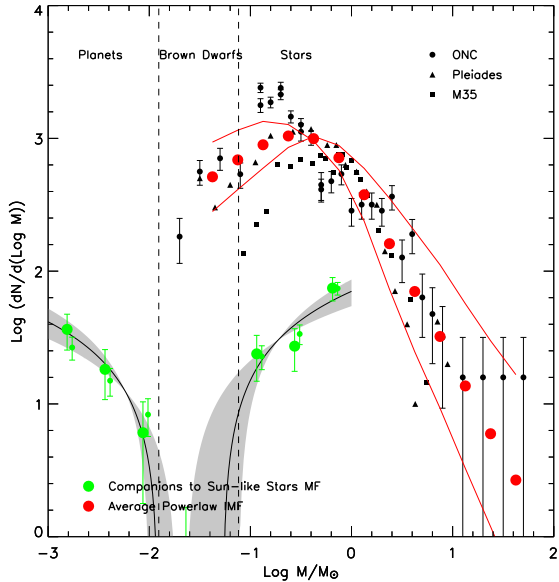}
\caption{The mass function of companions to Sun-like stars (lower left) compared to the initial mass function (IMF) of 
cluster stars (upper right). Our mass function of the companions to Sun-like stars is shown by the green dots (bigger dots 
are the $d < 25$ pc sample, smaller dots are the $d < 50$ pc sample). The linear slopes we fit to the data in Fig. 
\ref{fig:mass_25} are also shown along with their error. Data for the number of stars and brown dwarfs in the Orion Nebula 
Cluster (ONC) (circles), Pleiades cluster (triangles) and M35 cluster (squares) come from \citet{Hillenbrand00, Slesnick04}, 
\citet{Moraux03} and \citet{Barrado01} respectively and are normalized such that they overlap for masses larger than 
$1 M_\odot$ where a single power-law slope applies. The absolute normalization of cluster stars is arbitrary, while the 
companion mass function is normalized to the IMF of the cluster stars by scaling the three companion points of stellar mass 
to be on average $\sim 7\%$ for $P < 5$ years (derived from the stellar multiplicity of \citet{Duquennoy91} discussed in 
Section 3, combined with our estimate that $11\%$ of Sun-like stars have stellar secondaries). The average power-law IMF 
derived from various values of the slope of the IMF quoted in the literature \citep{Hillenbrand03} is shown as larger red 
dots along with two thin red lines showing the root-mean-square error. If the turn down in the number of brown dwarfs of 
the IMF is due to a selection effect because it is hard to detect brown dwarfs, then the two distributions are even more 
different from each other. For clarity the smaller green dots are shifted slightly to the right.
}
\label{fig:logN_logm}
\end{figure}

\begin{figure}[!t]
\plotone{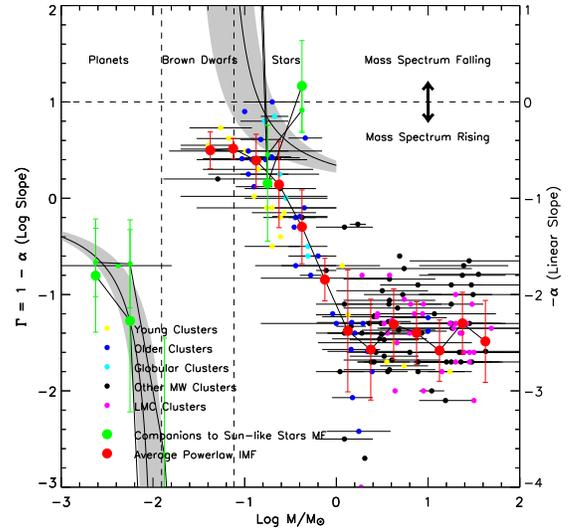}
\caption{
The initial mass function (IMF) for clusters represented by a series of power-law slopes \citep{Hillenbrand03}. Each point 
represents the power-law slope claimed to apply within the mass range indicated by the horizontal lines. Although the IMF 
is represented by a series of power-laws, the IMF is not a power-law for masses less than $1 M_\odot$ where the slope 
continually changes. The green dots show the slope of the companion mass function to Sun-like stars between the bins of 
Figs. \ref{fig:mass_25} \& \ref{fig:mass_50} with the larger and smaller dots respectively. The linear fits to the data 
in Fig. \ref{fig:mass_25} and their associated error are shown by the curves inside the grey regions. The power-law fit of 
\citet{Lineweaver03} (shown as the green dot with a horizontal line indicating the range over which the slope applies) is 
consistent with these fits. The larger red dots with error bars represent the average power-law IMF with a root-mean-square 
error. $\Gamma$ and $- \alpha$ are the respective logarithmic and linear slopes of the mass function. The logarithmic mass 
power-law distribution is $dN/dlogM \propto M^{\Gamma}$ and the linear mass power-law distribution is 
$dN/dM \propto M^{- \alpha}$ where $\Gamma = 1 - \alpha$. The errors on the fits of Fig. \ref{fig:mass_25} get smaller at 
$M \sim 10^{-3}\: M_{\odot}$ and $M \sim 1 \; M_{\odot}$ since as $log (M/M_{\odot})$ tends to $\pm \infty$, $\Gamma$ 
tends to 0. This can also be seen in Fig. \ref{fig:logN_logm} where the slopes of the upper and lower contours become 
increasingly similar.
}
\label{fig:IMF}
\end{figure}


\section{Comparison with the Initial Mass Function}

Brown dwarfs found as free-floating objects in the solar neighbourhood and as members of young star clusters have been 
used to extend the initial mass function (IMF) well into the brown dwarf regime. Comparing the mass function of our 
sample of close-orbiting companions of Sun-like stars to the IMF of single stars indicates how the environment of a 
host affects stellar and brown dwarf formation and/or migration. Here we quantify how different the companion mass 
function is from the IMF \citep{Halbwachs00}. 

The galactic IMF appears to be remarkably universal and independent of environment and metallicity with the possible 
exception of the substellar mass regime. A weak empirical trend with metallicity is suggested for very low mass stars and 
brown dwarfs where more metal rich environments may be producing relatively more low mass objects \citep{Kroupa02}. This 
is consistent with an extrapolation up in mass from the trend found in exoplanet hosts. The IMF is often represented as a 
power-law, although this only appears to be accurate for stars with masses above $\sim 1 M_\odot$ \citep{Hillenbrand03}. 
The stellar IMF slope gets flatter towards lower masses and extends smoothly and continously into the substellar mass 
regime where it appears to turn over.

Free floating brown dwarfs may be formed either as ejected stellar embryos or from low mass protostellar cores that have 
lost their accretion envelopes due to photo-evaporation from the chance proximity of a nearby massive star \citep{Kroupa03}. 
This hypothesis may explain their occurence in relatively rich star clusters such as the Orion Nebula cluster and their 
virtual absence in pre-main sequence stellar groups such as Taurus-Auriga.

In Figs. \ref{fig:logN_logm} \& \ref{fig:IMF} we compare the mass function of companions to Sun-like stars with the IMF of 
cluster stars. The mass function for companions to Sun-like stars is shown by the green dots from Figs. \ref{fig:mass_25} 
and \ref{fig:mass_50} (bigger dots are the $d < 25$ pc sample and smaller dots are the $d < 50$ pc sample). The linear 
slopes from Fig. \ref{fig:mass_25} and their one sigma confidence region are also shown. Between 
$log (M/M_{\odot}) \approx -1.0$ and $-0.5$ ($0.1 M_{\odot} < M < 0.3 M_{\odot}$) the slopes are similar. However, above
$0.3 M_{\odot}$ and below $0.1 M_{\odot}$ the slopes become inconsistent. Above $0.3 M_{\odot}$ the slopes, while of 
similar magnitude are of opposite sign and below $0.1 M_{\odot}$ the companion slope is much steeper than the IMF slope. 
The IMF for young clusters (yellow dots) is statistically indistinguishable from that of older stars (blue dots) and 
follows the average IMF.


\section{Summary and Discussion}

We analyse the close-orbitting ($P < 5$ years) planetary, brown dwarf and stellar companions to Sun-like stars to help 
constrain their formation and migration scenarios. We use the same sample to extract the relative numbers of planetary, 
brown dwarf and stellar companions and verify the existence of a brown dwarf desert. Both planetary and stellar companions 
reduce in number towards the brown dwarf mass range. We fit the companion mass function over the range that we analyse 
($0.001 < M/M_\odot \lsim 1.0$) by two separate straight lines fit separately to the planetary and stellar data points. 
The straight lines intersect in the brown dwarf regime, at $M = 31 \:^{+25}_{-18} \: M_{Jup}$. This result is robust to 
the declination asymmetry correction (Table \ref{table:minima}).

The period distribution of close-orbitting ($P < 5$ years) companion stars is different from that of the planetary 
companions. The close-in stellar companions are fairly evenly distributed over $logP$ with planets tending to be clumped 
towards higher periods. We compare the companion mass function to the IMF for bodies in the brown dwarf and stellar 
regime. We find that starting at $1 \; M_{\odot}$ and decreasing in mass, stellar companions continue to reduce in number 
into the brown dwarf regime, while cluster stars increase in number before reaching a maximum just before the brown dwarf 
regime (Fig. 13). This leads to a difference of at least 1.5 orders of magnitude between the much larger number of brown 
dwarfs found in clusters to those found as close-orbitting companions to Sun-like stars.

The period distribution of close-orbiting companions may be more a result of post-formation migration and gravitational 
jostling than representive of the relative number of companions that are formed at a specific distance from their hosts. 
The companion mass distribution is more fundamental than the period distribution and should provide better constraints on 
formation models, but our ability to sample the mass distribution is only for $P < 5$ years.

We show in Figs. \ref{fig:mass_host} and \ref{fig:mass_host_50} that lower mass hosts have more stellar companions and 
fewer giant planet companions while higher mass hosts have fewer stellar companions but more giant planet companions.
The brown dwarf desert is generally thought to exist at close separations $\lsim$ 3 AU (or equivalently $P \leq 5$ years) 
\citep{Marcy00} but may disappear at wider separations. \citet{Gizis01} suggests that at very large separations 
($> 1000$ AU) brown dwarf companions may be more common. However, \citet{McCarthy04} in their observation of 280 GKM stars 
find only 1 brown dwarf between 75 and 1200 AU. \citet{Gizis03} reports that $15\% \pm 5 \%$ of M/L dwarfs are brown dwarf 
binaries with separations in the range $1.6 - 16$ AU. This falls to $5\% \pm 3 \%$ of M/L dwarfs with separations less 
than 1.6 AU and none with separations greater than 16 AU. This differs greatly from the brown dwarfs orbiting Sun-like 
stars but is consistent with our host/minimum-companion-mass relationship, i.e., we expect no short period brown dwarf 
desert around M or L type stars.

Three systems containing both a companion with a minimum mass in the planetary regime and a companion with a minimum mass 
in the brown dwarf regime are known - HD 168443 \citep{Marcy01}, HD 202206 \citep{Udry02, Correia04} and GJ 86 
\citep{Queloz00, Els01}. Our analysis suggests that both the $M sin(i)$-brown dwarfs orbiting HD 168443 and HD 202206 are 
probably stars (see Section 2.2 for our false positive brown dwarf correction). If the $M sin(i)$-planetary companions 
in these 2 systems are coplanar with the larger companions then these ``planets" may be brown dwarfs or even stars. GJ 86 
contains a possible brown dwarf detected orbiting at $\sim 20$ AU ($P > 5$ years) and so was not part of our analysis. 
However this does suggest that systems containing stars, brown dwarfs and planets may be possible.

We find that approximately $16\%$ of Sun-like stars have a close companion more massive than Jupiter. Of these $16\%$, 
$11\% \pm 3\%$ are stellar, $<1\%$ are brown dwarf and $5\% \pm 2\%$ are planetary companions (Table \ref{table:fraction}). 
Although \citet{Lineweaver03} show that the fraction of Sun-like stars with planets is greater than $25\%$, this is for 
target stars that have been monitored the longest ($\sim 15$ years) and at optimum conditions (stars with low-level chromospheric 
activity or slow rotation) using the high precision Doppler method. When we limit the analysis of \citet{Lineweaver03} to 
planetary companions with periods of less than 5 years and masses larger than Jupiter, we find the same value that we 
calculate here. When we split our sample of companions into those with hosts above and below $1 M_\odot$, we find that for 
the lower mass hosts: $11.8\%$ have stellar, $<1\%$ have brown dwarf and $4.2\%$ have planetary companions and that for the 
higher mass hosts: $9.4\%$ have stellar, $<1\%$ have brown dwarf and $6.6\%$ have planetary companions respectively 
(Table \ref{table:fraction}). More massive hosts have more planets and fewer stellar companions than less massive hosts. 
These are marginal results but are seen in both the 25 and 50 pc samples.

The constraints that we have identified for the companions to Sun-like stars indicate that close orbitting brown dwarfs 
are very rare. The fact that there is a close-orbitting brown dwarf desert but no free floating brown dwarf desert suggests 
that post-collapse migration mechanisms may be responsible for this relative dearth of observable brown dwarfs rather than 
some intrinsic minimum in fragmentation and gravitational collapse in the brown dwarf mass regime \citep{Ida04}. Whatever 
migration mechanism is responsible for putting hot Jupiters in close orbits, its effectiveness may depend on the mass ratio 
of the object to the disk mass. Since there is evidence that disk mass is correlated to host mass, the migratory mechanism 
may be correlated to host mass, as proposed by \citet{Armitage02}.


\section{Acknowledgements}

We would like to thank Christian Perrier for providing us with the Elodie exoplanet target list, Stephane Udry for 
addititional information on the construction of the Coralie exoplanet target list and Lynne Hillenbrand for sharing
her data collected from the literature on the power-law IMF fits to various stellar clusters. This research has 
made use of the SIMBAD database, operated at CDS, Strasbourg, France. This research has made use of the Washington 
Double Star Catalog maintained at the U.S. Naval Observatory.


\section{Appendix: Companion Mass Estimates}

The Doppler method for companion detection cannot give us the mass of a companion without some additional astrometric or 
visual solution for the system or by making certain assumptions about the unknown inclination except in the case where 
a host star and its stellar companion have approximately equal masses and a double-lined solution is available. Thus to 
find the companion mass $M_2$ that induces a radial velocity $K_1$ in a host star of mass $M_1$ we use (see Heacox 1999)

\begin{equation}
K_1 = {(\frac{2 \pi G}{P})}^{1/3} \frac{M_2 sin(i)}{(M_1 + M_2)^{2/3}} \frac{1}{{(1 - e^2)}^{1/2}}
\end{equation}

This equation can be expressed in terms of the mass function $f(m)$

\begin{equation}
f(m) = \frac{M_2^3 sin^3(i)}{(M_1 + M_2)^2} = \frac{P K_1^3 (1 - e^2)^{3/2}}{2 \pi G}
\end{equation}

Eq. 3 can then be expressed in terms of a cubic equation in the mass ratio $q = M_2 / M_1$, where $Y = f(m) / M_1$.

\begin{equation}
q^3 sin^3(i) - Yq^2 - 2Yq - Y = 0
\end{equation}

For planets ($M_1 >> M_2$) we can simplify Eq. 2 and directly solve for $M_2 sin(i)$ but this is not true for larger 
mass companions such as brown dwarfs and stars. We use \citet{Cox00} to relate host mass to spectral type. When a 
double-lined solution is available, the companion mass can be found from $q = M_2 / M_1 = K_1 / K_2$. 

For all single-lined Doppler solutions, where the inclination $i$ of a companion's orbit is unknown (no astrometric or 
visual solution), we assume a random distribution $P(i)$ for the orientation of the inclination with respect to our 
line of sight,

\begin{equation}
P(i) di = sin(i) di
\end{equation}

From this we can find probability distributions for $sin(i)$ and $sin^3(i)$. \citet{Heacox95} and others suggest using 
either the Richardson-Lucy or Mazeh-Goldberg algorithms to approximate the inclination distribution. However, 
\citet{Hogeveen91} and \citet{Trimble90} argue that for low number statistics, the simple mean method produces similar 
results to the more complicated methods. We have large bin sizes and small number statistics, hence we use this method. 
The average values of the $sin(i)$ and $sin^3(i)$ distributions assuming a random inclination are $<sin(i)> = 0.785$ and 
$<sin^3(i)> = 0.589$, which are used to estimate the mass for planets and other larger single-lined spectroscopic binaries 
respectively. For example, in Fig. \ref{fig:Binary_m_P}, of the 198 mass estimates in the 50 pc sample, 53 ($27\%$) come 
from visual double-lined Doppler solutions, 6 ($3\%$) come from infrared double-lined Doppler solutions \citep{Mazeh03},
18 ($9\%$) come from knowing the inclination (astrometric or visual solution also available for system), 10 ($5\%$) come 
from assuming that Doppler brown dwarf candidates have low inclinations, 55 ($28\%$) come from assuming $<sin(i)> = 0.785$ 
and 56 ($28\%$) from assuming $<sin^3(i)> = 0.589$.




\begin{thebibliography}{}

\bibitem[Armitage \& Bonnell(2002)]{Armitage02}
Armitage, P.J. \& Bonnell, I.A., 2002,
`The Brown Dwarf Desert as a Consequence of Orbital Migration',
MNRAS, 330:L11

\bibitem[Barrado y Navascues \etal(2001)]{Barrado01}
Barrado y Navascues, D., Stauffer, J.R., Bouvier, J. \& Martin, E.L., 2001,
`From the Top to the Bottom of the Main Sequence: A Complete Mass Function of the Young Open Cluster M35',
\apj, 546:1006-1018

\bibitem[Bate(2000)]{Bate00}
Bate, M.R., 2000,
`Predicting the Properties of Binary Stellar Systems: The Evolution of Accreting Protobinary Systems',
MNRAS, 314:33-53

\bibitem[Boss(2002)]{Boss02}
Boss, A.P., 2002,
`Evolution of the Solar Nebula V: Disk Instabilities with Varied Thermodynamics',
\apj, 576:462-472

\bibitem[Burrows(1997)]{Burrows97}
Burrows, A., Marley, M., Hubbard, W.B., Lunine, J.I., Guillot, T., Saumon, D., Freedman, R., Sudarsky, D.; \& Sharp, C., 1997,
`A Nongray Theory of Extrasolar Giant Planets and Brown Dwarfs',
\apj, 491:856

\bibitem[Correia \etal(2004)]{Correia04}
Correia, A.C.M., Udry, S., Mayor, M., Laskar, J., Naef, D., Pepe, F., Queloz, D. \& Santos, N.C., 2004, 
`The CORALIE survey for southern extra-solar planets XIII. A pair of planets around HD 202206 or a circumbinary planet?',
A\&A, 440, 751-758

\bibitem[Cox(2000)]{Cox00}
Cox, A.N., 2000,
`Allen's Astrophysical Quantities',
AIP Press, 4th Edition

\bibitem[Duquennoy \& Mayor(1991)]{Duquennoy91}
Duquennoy, A. \& Mayor, M., 1991,
`Multiplicity among Solar-type Stars in the Solar Neighbourhood II',
A\&A, 248:485-524

\bibitem[Els \etal(2001)]{Els01}
Els, S.G., Sterzik, M.F., Marchis, F., Pantin, E., Endl, M. \& Kürster, M., 2001,
`A Second Substellar Companion in the Gliese 86 System. A Brown Dwarf in an Extrasolar Planetary System',
A\&A, 370:L1-L4

\bibitem[Endl \etal(2004)]{Endl04}
Endl, M., Hatzes, A.P., Cochran, W.D., McArthur, B., Allende Prieto, C., Paulson, D.B., Guenther, E. \& Bedalov, A., 2004,
`HD 137510: An Oasis in the Brown Dwarf Desert',
\apj, 611:1121-1124

\bibitem[Hipparcos Catalog(1997)]{HIP}
ESA, The Hipparcos and Tycho Catalogues, 1997, ESA SP-1200
http://astro.estec.esa.nl/hipparcos/

\bibitem[Fischer \etal(1999)]{Fischer99}
Fischer, D.A., Marcy, G.W., Butler, P.R., Vogt, S.S. \& Apps, K., 1999,
`Planetary Companions around Two Solar-Type Stars: HD 195019 and HD 217107',
PASP, 111:50-56

\bibitem[Gizis \etal(2001)]{Gizis01}
Gizis, J.E., Kirkpatrick, J.D., Burgasser, A., Reid, I.N., Monet, D.G., Liebert, J. \& Wilson, J.C., 2001,
`Substellar Companions to Main-Sequence Stars: No Brown Dwarf Desert at Wide Separations',
\apj, 551:L163-L166

\bibitem[Gizis \etal(2003)]{Gizis03}
Gizis, J.E., Reid, I.N., Knapp, G.R., Liebert, J., Kirkpatrick, J.D., Koerner, D.W. \& Burgasser, A.J., 2003,
`Hubble Space Telescope Observations of Binary Very Low Mass Stars and Brown Dwarfs',
AJ, 125:3302-3310

\bibitem[Hartkopf \& Mason(2004)]{WDS6}
Hartkopf, W.I. \& Mason, B.D., 2004,
`Sixth Catalog of Orbits of Visual Binary Stars',
http://ad.usno.navy.mil/wds/orb6.html

\bibitem[Halbwachs \etal(2000)]{Halbwachs00}
Halbwachs, J.L., Arenou, F., Mayor, M., Udry, S. \& Queloz, D., 2000,
`Exploring the Brown Dwarf Desert with Hipparcos',
A\&A, 355:581-594

\bibitem[Halbwachs \etal(2003)]{Halbwachs03}
Halbwachs, J.L., Mayor, M., Udry, S. \& Arenou, F., 2003,
`Multiplicity among Solar-type Stars III',
A\&A, 397:159-175

\bibitem[Heacox(1995)]{Heacox95}
Heacox, W.D., 1995,
`On the Mass Ratio Distribution of Single-Lined Spectroscopic Binaries',
AJ, 109, 6:2670-2679

\bibitem[Heacox(1999)]{Heacox99}
Heacox, W.D., 1999,
`On the Nature of Low-Mass Companions to Solar-like Stars',
\apj, 526:928-936

\bibitem[Hillenbrand(2003)]{Hillenbrand03}
Hillenbrand, L.A., 2003,
`The Mass Function of Newly Formed Stars',
astro-ph/0312187

\bibitem[Hillenbrand \& Carpenter(2000)]{Hillenbrand00}
Hillenbrand, L.A. \& Carpenter, J.M., 2000,
`Constraints on the Stellar/Substellar Mass Function in the Inner Orion Nebula Cluster',
\apj, 540:236-254

\bibitem[Hogeveen (1991)]{Hogeveen91}
Hogeveen, S.J., 1991,
Ph.D. Thesis, University of Illinois, Urbana 

\bibitem[Ida \& Lin (2004)]{Ida04}
Ida, S. \& Lin, D.N.C., 2004,
`Toward a Deterministic model of planetary formation. I. A desert in the mass and semimajor axis distribution of extrasolar planets',
\apj, 604:388-413

\bibitem[Jiang, Laughlin \& Lin(2004)]{Jiang04}
Jiang, I.-G., Laughlin, G. \& Lin, D.N.C., 2004,
`On the Formation of Brown Dwarfs',
\apj, 127:455-459

\bibitem[Jones \etal(2002)]{Jones02}
Jones, H.R.A., Butler, P.R., Marcy, G.W., Tinney, C.G., Penny, A.J., McCarthy, C. \& Carter, B.D., 2002,
`Extra-solar planets around HD 196050, HD 216437 and HD 160691',
\mnras, 337:1170-1178

\bibitem[Konacki \etal(2004)]{Konacki04}
Konacki, M., Torres, G., Sasselov, D.D., Pietrzynski, G., Udalski, A., Jha, S., Ruiz, M.T., Gieren, W. \& Minniti, D., 2004,
`The Transiting Extrasolar Giant Planet Around the Star OGLE-TR-113',
\apj, 609:L37-L40

\bibitem[Kroupa(2002)]{Kroupa02}
Kroupa, P., 2002,
`The Initial Mass Function of Stars: Evidence for Uniformity in Variable Systems',
Science, 295:82-91

\bibitem[Kroupa \& Bouvier(2003)]{Kroupa03}
Kroupa, P. \& Bouvier, J., 2003,
`On the Origin of Brown Dwarfs and Free-Floating Planetary-Mass Objects',
MNRAS, 346:369-380

\bibitem[Larson(2003)]{Larson03}
Larson, R.B., 2003,
`The Physics of Star Formation',
astro-ph/0306596

\bibitem[Lineweaver \etal (2003)]{LGH03}
Lineweaver, C.H., Grether, D. \& Hidas, M. 2003,
`What can exoplanets tell us about our Solar System?'
in the proceedings `Scientific Frontiers in Research on Extrasolar Planets',
ASP Conf. Ser. Vol. 294, edt Deming, D. \& Seager, S., p 161, astro-ph/0209382

\bibitem[Lineweaver \& Grether(2003)]{Lineweaver03}
Lineweaver, C.H. \& Grether, D., 2003,
`What Fraction of Sun-Like Stars have Planets?',
\apj, 598:1350-1360

\bibitem[Marcy \& Butler(2000)]{Marcy00}
Marcy, G.W. \& Butler, P.R., 2000,
`Planets Orbiting Other Suns',
PASP, 112:137-140 

\bibitem[Marcy \etal(2001)]{Marcy01}
Marcy, G.W., Butler, P.R., Vogt, S.S., Liu, M.C., Laughlin, G., Apps, K., Graham, J.R., Lloyd, J., Luhman, K.L. \& Jayawardhana, R., 2001,
`Two Substellar Companions Orbiting HD 168443',
\apj, 555:418-425

\bibitem[Matzner \& Levin(2004)]{Matzner04}
Matzner, C.D. \& Levin, Y., 2004,
`Low-Mass Star Formation: Initial Conditions, Disk Instabilities and the Brown Dwarf Desert',
astro-ph/0408525 

\bibitem[Mazeh \etal(2003)]{Mazeh03}
Mazeh, T., Simon, M., Prato, L., Markus, B. \& Zucker, S., 2003,
`The Mass Ratio Distribution in Main-Sequence Spectroscopic Binaries Measured by IR Spectroscopy',
\apj, 599:1344-1356

\bibitem[McCarthy \& Zuckerman(2004)]{McCarthy04}
McCarthy, C. \& Zuckerman, B., 2004,
`The Brown Dwarf Desert at 75-1200 AU',
AJ, 127:2871-2884

\bibitem[Moraux \etal(2003)]{Moraux03}
Moraux, E., Bouvier, J., Stauffer, J.R. \& Cuillandre, J.C., 2003,
`Brown Dwarfs in the Pleiades Cluster: Clues to the Substellar Mass Function',
A\&A, 400:891-902

\bibitem[Nidever \etal(2002)]{Nidever02}
Nidever, D.L., Marcy, G.W., Butler, P.R., Fischer, D.A. \& Vogt, S.S., 2002,
`Radial Velocities for 889 Late-type Stars',
ApJSS, 141:503-522

\bibitem[Pont \etal(2004)]{Pont04}
Pont, F., Bouchy, F., Queloz, D., Santos, N.C., Mayor, M. \& Udry, S., 2004,
`The Missing Link: A 4-day Period Transiting Exoplanet around OGLE-TR-111',
A\&A, 426:L15-L18

\bibitem[Pourbaix \etal(2004)]{SB9}
Pourbaix D., Tokovinin A.A., Batten A.H., Fekel F.C., Hartkopf W.I., Levato H., Morrell N.I., Torres G., Udry S., 2004,
`SB9: The Ninth Catalogue of Spectroscopic Binary Orbits',
A\&A, 424:727-732

\bibitem[Queloz \etal(2000)]{Queloz00}
Queloz, D., Mayor, M., Weber, L., Blécha, A., Burnet, M., Confino, B., Naef, D., Pepe, F., Santos, N. \& Udry, S., 2000,
`The CORALIE Survey for Southern Extra-Solar Planets. I. A planet Orbiting the Star Gliese 86',
A\&A, 354:99-102

\bibitem[Reid(2002)]{Reid02}
Reid, I.N., 2002,
`On the Nature of Stars with Planets',
PASP, 114:306-329

\bibitem[Rice \etal(2003)]{Rice03}
Rice, W.K.M., Armitage, P.J., Bonnell, I.A., Bate, M.R., Jeffers, S.V. \& Vine, S.G., 2003,
`Substellar Companions and Isolated Planetary Mass Objects from Protostellar Disk Fragmentation',
MNRAS, 346:L36-L40

\bibitem[Saar \etal(1998)]{Saar98}
Saar, S.H., Butler, P.R. \& Marcy, G.W., 1998,
`Magnetic Activity Related Radial Velocity Variations in Cool Stars: First Results from Lick Extrasolar Planet Survey',
\apj, 498:L153-L157

\bibitem[Schneider(2005)]{EPC}
Schneider, J., 2005,
`Extrasolar Planets Catalog',
http://www.obspm.fr/encycl/catalog.html

\bibitem[Slesnick \etal(2004)]{Slesnick04}
Slesnick, C.L., Hillenbrand, L.A. \& Carpenter, J.M., 2004,
`The Spectroscopically Determined Substellar Mass Function of the Orion Nebula Cluster',
\apj, 610:1045-1063

\bibitem[Tinney \etal(2001)]{Tinney01}
Tinney, C.G., Butler, P.R., Marcy, G.W., Jones, H.R.A., Vogt, S.S., Apps, K. \& Henry, G.W., 2001,
`First Results from the Anglo-Australian Planet Search',
\apj, 551:507-511

\bibitem[Trimble(1990)]{Trimble90}
Trimble, V., 1990,
`The Distributions of Binary System Mass Ratios: A Less Biased Sample',
MNRAS, 242:79-87

\bibitem[Udry \etal(2002)]{Udry02}
Udry, S., Mayor, M., Naef, D., Pepe, F., Queloz, D., Santos, N.C. \& Burnet, M., 2002,
`The CORALIE survey for southern extra-solar planets VIII. The very low-mass companions of HD 141937, HD 162020, 
HD 168443, HD 202206: Brown dwarfs or superplanets?',
A\&A, 390, 267-279

\bibitem[Udry \etal(2003)]{Udry03}
Udry, S., Mayor, M. \& Santos, N.C., 2003,
`Statistical Properties of Exoplanets I: The Period Distribution - Constraints for the Migration Scenario',
A\&A, 407:369-376

\bibitem[Vogt \etal(2002)]{Vogt02}
Vogt, S.S., Butler, P.R., Marcy, G.W., Fischer, D.A., Pourbaix, D., Apps, K. \& Laughlin, G., 2002,
`Ten Low-Mass Companions from the Keck Precision Velocity Survey',
\apj, 568:352-362

\bibitem[Wright \etal(2004)]{Wright04}
Wright, J.T., Marcy, G.W., Butler, P.R. \& Vogt, S.S., 2004,
`Chromospheric Ca II Emission in Nearby F, G, K and M stars',
ApJSS, 152:261-295

\bibitem[Zucker \& Mazeh(2001a)]{Zucker01a}
Zucker, S. \& Mazeh, T., 2001a,
`Analysis of the Hipparcos Observations of the Extrasolar Planets and the Brown Dwarf Candidates',
\apj, 562:549-557

\bibitem[Zucker \& Mazeh(2001b)]{Zucker01b}
Zucker, S. \& Mazeh, T., 2001b,
`Derivation of the Mass Distribution of Extrasolar Planets with Maxlima, A Maximum Likelihood Algorithm',
\apj, 562:1038-1044

\end{thebibliography}
\end{document}